\documentclass[preprint,showpacs]{article}
\usepackage[centertags]{amsmath}
\usepackage{amsfonts}
\usepackage{amssymb}
\usepackage{amsthm} 
\usepackage{newlfont}
\usepackage{epsfig}
\usepackage{amscd}
\usepackage{graphicx}
\usepackage{epsfig}
\usepackage{color}
\usepackage{amscd}
\newcommand{\NN}{{\mathbb N}}

\newcommand{\beq}{\begin{equation}}
\newcommand{\eeq}{\end{equation}}
\newcommand{\ba}{\begin{array}}
\newcommand{\ea}{\end{array}}
\newcommand{\bea}{\begin{eqnarray}}
\newcommand{\eea}{\end{eqnarray}}
\begin{document}

\begin{center}
{\large \sc \bf {
Dynamical storage of quantum states
}}

\vskip 15pt

{\large 
E.B.Fel'dman and A.I.~Zenchuk 
}

\vskip 8pt

{\it Institute of Problems of Chemical Physics, RAS,
Chernogolovka, Moscow reg., 142432, Russia},\\

\end{center}


\begin{abstract}
We consider a dynamical method of storage of quantum states based on the spin-1/2 systems with the dipole-dipole interactions  in a strong external  magnetic field { supplemented with the special time-reversion procedure}. The stored information can be extracted at certain time instants. 
 \end{abstract}


\section{Introduction}
\label{Section:Introduction}
\label{Section:switching}
The storage of quantum states is an attractive area of quantum information processing. A set of proposed protocols  is based on  { various}  optical effects in the atomic 
systems \cite{LST,BSARST}. Among others we mention the control reversible inhomogeneous broadening \cite{MK}, an off-resonant Raman memory \cite{KMP,MKMP}, a silenced echo quantum memory \cite{Moiseev,DBLCG}. All these protocols are based on the storage of the information using excited energy levels and then extracting this information at the desired time instant.  

Here we consider another principle of memory organization which we call the dynamical storage of a state. It is based on the reversion of a quantum evolution and complete recovering of the original state provided the absence of interaction with an environment. { The possible candidates for implementation of our protocol are calcium fluorapatite, calcium hydroxyapatite  and optical lattices. The latter are most promising due to their flexibility in preparing required  coupling constants and local magnetic field}.  

{The simplest way  achieving} the  reversion of spin-evolution is based on  the special { tool 
(sample rotation or series of  magnetic pulses \cite{RPW}) } which provides the periodically  {  switching  sign of the Hamiltonian} governing the spin dynamics: from $H$ to $-\alpha H$,
$\alpha>0$. Thus, if the state evolves under the Hamiltonian 
 $H$ during the time interval  $\tau_1$ and then it evolves under $-\alpha H$ during the time interval $\tau_2$, then the overall evolution over the period $T=\tau_1+\tau_2$ is described by the  evolution operator $V$,
 \begin{eqnarray}\label{V}
 V=e^{i\alpha H\tau_2} e^{-iH\tau_1},
 \end{eqnarray}
so that
 \begin{eqnarray}\label{T}
 \rho(T)= V  \rho(0) V^+,
 \end{eqnarray}
 where $\rho(0)$ is the initial state.
 Let the time intervals $\tau_1$ and $\tau_2$  be related as
 \begin{eqnarray}\label{tau}
 \tau_1 = \alpha \tau_2. 
 \end{eqnarray}
 Then $V$ equals the identity operator $I$, therefore eq.(\ref{T}) yields
\begin{eqnarray}
\rho(T) \equiv \rho(0).
\end{eqnarray}
Thus, in the ideal case, there is a perfect initial state recovering. 
Obviously,
\begin{eqnarray}
\rho(nT) = V^n  \rho(0) (V^+)^n \equiv \rho(0). 
\end{eqnarray}
Therefore, if we keep switching the Hamiltonian, the system returns to the initial state at the discrete time instants 
$t_n=nT$, $n\in \NN$.

{The time reversion using the above switching  Hamiltonians can be implemented, { for instance, } in the line systems as shown below in Sec.\ref{Section:line}. In general, the time reversion requires more complicated switching of Hamiltonians and can be realized by the method of the average Hamiltonian theory \cite{HW}. Two such methods of time-reversion are discussed in Sec.\ref{Section:TR1}. Examples of implementation 
of { the state-storage protocol  to bipartite systems } are considered in Sec.\ref{Section:application}. Concluding remarks are 
given in Sec.\ref{Section:conclusion}.}

\section{Time reversion protocol }
\label{Section:TR1}

{
The problem can be stated as follows. Let $\rho(0)$ be the initial state of some multi-qubit quantum system whose dynamics is described by  some evolution operator $V(t)$,  $\rho(t)=V(t) \rho(0) V^+(t)$. Our purpose is to handle the interaction Hamiltonian in a way such that, at some time instant $T$, we have 
$\rho(T)=V(T) \rho(0) V^+(T) \equiv \rho(0)$, i.e., $V(T)$ is equivalent to the identity operator $I$: $V(T)\equiv I$.  
Two protocols of constructing such evolution operators are proposed in Secs. \ref{Section:rev1} and \ref{Section:rev2}.}
The 
first of them  is based { on the suitable rotation of a sample with a spin system relative to the stationary magnetic field}, while the second one  is based on the certain  unitary transformations { of a  spin system} (irradiation by the  periodic sequence of magnetic pulses)   without {rotation of a sample}. 

In both cases we start with the anisotropic  Hamiltonian of the dipole-dipole interaction   governing the spin dynamics in the  strong $z$-directed magnetic field  
($H_{dz}$-Hamiltonian) which reads in the frequency units  as { (we use the rotating reference frame)}
\begin{eqnarray}\label{HXY}
&&
H\equiv H_{dz}=\sum_{j>i} D_{ij}( 2I_{zi} I_{zj} - I_{xi} I_{xj}-  I_{yi} I_{yj}) ,\\\label{DXY}
&&
D_{ij}= \frac{\gamma^2 \hbar}{2 r_{ij}^3} (1-3 \cos^2 \theta_{ij}).
\end{eqnarray} 
Here $\gamma$ is the gyromagnetic ratio { of spin particles}, $r_{ij}$ is the distance between the $i$th and $j$th spins, $\theta_{ij}$ is the angle between the vector $r_{ij}$ and direction of the magnetic field, and $I_{\alpha i}$ is the projection of the { $i$th spin} angular momentum  on the axis $\alpha$, $\alpha=x,y,z$. 

\subsection{Reversion via sample orientation} 
\label{Section:rev1}
We see that the coupling constants $D_{ij}$ depend on the spin-system orientation  through the angles $\theta_{ij}$. Therefore  the protocol of   time-reversion depends on the space-dimensionality of the spin system.  We propose such protocols  for  one-   and two-dimensional  spin systems.  The only requirement to the magnetic field ${\cal{H}}$ is $\omega \equiv \gamma {\cal{H}} \gg \omega_{loc}$. 
{ Emphasize that the field-orientation in this  subsection is provided by the proper rotation of the sample (spin system) rather than  by the rotation of the magnetic field. This fact allows us to use the Hamiltonian (\ref{HXY}) during the whole experiment. 
}

\subsubsection{One-dimensional system (chain)}
\label{Section:line}
In  this case all $\theta_{ij}$ equal each other: $\theta_{ij}\equiv \theta$.
We   consider  Hamiltonian $H$  as a function of $\theta$ and note that 
\begin{eqnarray}\label{HH}
H_{dz}|_{\theta=0}\equiv H_{dz}(0)=-2 H_{dz}(\frac{\pi}{2})\equiv -2 H_{dz}|_{\theta=\frac{\pi}{2}}.
\end{eqnarray}
Therefore, we can create the evolution  operator $V$  of form (\ref{V},\ref{tau}) with $H=H_{dz}(\frac{\pi}{2})$, 
$\alpha=2$,  $\tau_1=2\tau_2$ and $T=3\tau_2$. 

\subsubsection{Two-dimensional spin-lattice}
\label{Section:plane}

 In the case of two-dimensional spin system the protocol of the time reversion { differs from that in Sec.\ref{Section:line} } and consists of three steps. { Each time we rotate the sample with the spin system to reach the required magnetic field orientation. }

\begin{enumerate}
\item\label{item1}
{ The magnetic field is in the plane of the spin system. The dipolar coupling constants read}
\begin{eqnarray}\label{D2}
 D^{1}_{ij}= \frac{\gamma^2 \hbar}{2 r_{ij}^3} (1-3 \cos^2 \theta_{ij}).
\end{eqnarray}
\item
{
The magnetic field remains in the plane of the spin system, its direction is perpendicular to that in  n.\ref{item1}. Then}
\begin{eqnarray}\label{D3}
 D^2_{ij}= \frac{\gamma^2\hbar}{2 r_{ij}^3} (-2+3 \cos^2 \theta_{ij}),
\end{eqnarray}
where $\theta_{ij}$ are angles used in (\ref{D2}).
\item
{ 
The  magnetic field is perpendicular to the plane of the  spin system. Then $\theta_{ij}=\frac{\pi}{2}$, and}
\begin{eqnarray}\label{D4}
 D^3_{ij}= \frac{\gamma^2 \hbar}{2 r_{ij}^3}.
\end{eqnarray}
\end{enumerate}
Obviously, 
\begin{eqnarray}
\label{D234}
D^{2}_{ij}+D^3_{ij}+D^1_{ij} = 0.
\end{eqnarray}
Thus the overall evolution over the period $T$ is defined by the operator (we denote $H_k$ the Hamiltonian 
(\ref{HXY}) associated with $D^k_{ij}$, $k=1,2,3$)
\begin{eqnarray}\label{ev210}
V=e^{-i H_1 \tau}e^{-i H_2 \tau} e^{-i H_3 \tau}.
\end{eqnarray}
Let the above   change of orientation of  the magnetic field { proceeds} periodically and the period $T=3 \tau$ is sufficiently small, i.e., $T\omega_{loc}\ll 1$, where $\omega_{loc}$ is the local dipolar field \cite{G}. This allows us to introduce a new time 
$\eta=m T$, $m\in \NN$,  and introduce the effective Hamiltonian via the average Hamiltonian theory \cite{HW}:
$H_{eff}=\frac{1}{3}(H_1+H_2+H_3)$. However, due to (\ref{D234}), $H_{eff}\equiv 0$. Therefore we have 
\begin{eqnarray}\label{ev21}
V(\eta)=V^m(T)=e^{-i H_{eff}\eta} \equiv I,
\end{eqnarray}
and the density matrix $\rho(m T)=\rho(0)$, which holds up to the terms of the order $(T \omega_{loc})^2$. 

\subsection{Time-reversion protocol via  unitary transformations}
\label{Section:rev2}
\begin{figure*}
\hspace*{-0.3cm}  \epsfig{file=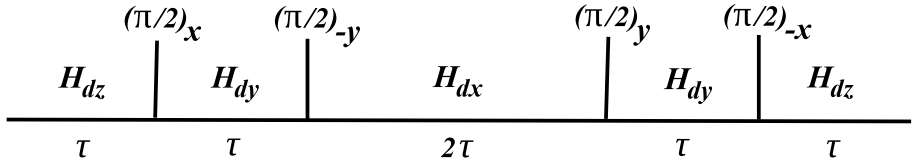,
  scale=0.5
   ,angle=0
}  
\caption{The periodic  pulse sequence  providing  effective Hamiltonian (\ref{Heff}).} 
\label{Fig:pulses}
\end{figure*}

{ Now we consider a spin system in the strong $z$-directed magnetic field ${\cal{H}}_z$ ($\gamma {\cal{H}}_z \gg \omega_{loc}$)), 
so that the system evolves under the secular part of  $H_{dz}$ Hamiltonian (\ref{HXY}) initially.
Our protocol is based on the irradiation of  a spin system by the  sequence of the resonance periodic 
 $(\frac{\pi}{2})_{\pm x}$ and
 $(\frac{\pi}{2})_{\pm y}$ pulses (with the magnetic field intensity ${\cal{H}}$ such that
 $\gamma {\cal{H}}_z\gg \gamma {\cal{H}} \gg \omega_{loc}$) shown in Fig.\ref{Fig:pulses}, 
 which is similar to that introduced in \cite{HW,RPW,PF}. } { We emphasize that, unlike the protocol in Sec.\ref{Section:rev1}, the sample with the  spin system
  is in a fixed position and we  control the direction of the  magnetic field in pulses.}

Due to the sequence of magnetic pulses, the Hamiltonians change each other during the single period $T=6 \tau$ as shown in 
Fig. \ref{Fig:pulses}, where 
\begin{eqnarray}\label{Hdy}
&&
H_{dy}=\sum_{j>i} D_{ij}(2I_{yi} I_{yj} -I_{xi} I_{xj}- I_{zi} I_{zj}),\\\label{Hdx}
&&
H_{dx}=\sum_{j>i} D_{ij}(2I_{xi} I_{xj}-I_{yi} I_{yj}- I_{zi} I_{zj}) .
\end{eqnarray} 
The overall evolution over the period $T $ is defined as follows
\begin{eqnarray}
V(T)= e^{-i\tau H_{dz}  } e^{-i \tau H_{dy} }
 e^{- 2 i  \tau H_{dx}  } e^{- i \tau H_{dy} } e^{-i \tau H_{dz} }. 
\end{eqnarray}
Now we repeat the sequence shown in Fig.\ref{Fig:pulses} $m$ times and introduce the time $\eta = m T$, $m\gg 1$. Then the evolution in time $\eta$ of the system can be described by the operator  
\begin{eqnarray}
V(\eta)\equiv V^m(T)=e^{-i H_{eff} \eta}
\end{eqnarray}
with
\begin{eqnarray}\label{Heff}
H_{eff} = \frac{1}{3} ( H_{dx} + H_{dy} + H_{dz}),
\end{eqnarray}
{derived via the average Hamiltonian theory \cite{HW}.}
However, the direct calculations show that 
\begin{eqnarray}\label{Hdxyz}
H_{dx}+H_{dy}+H_{dz}=0.
\end{eqnarray}
Therefore,
\begin{eqnarray}
H_{eff} \equiv 0,
\end{eqnarray}
and the evolution operator becomes the identity operator $I$:
\begin{eqnarray}\label{ev}
V(\eta) =  e^{-i H_{eff} \eta} \equiv I.
\end{eqnarray}

\section{State-storage protocol in bipartite systems}
\label{Section:application}

\subsection{Frozen state of subsystem}
\label{Section:frozen}

{ Now we show how the state-storage protocol described in Sec.\ref{Section:TR1} can be implemented in simple quantum processes. }

{ The first obstacle we meet is the organizing the proper  interaction between  the dynamical memory (a subsystem used to store a multi-qubit state) and 
the transmission line supplying the state needed to store. Formally, the problem can be stated as follows.}

Let us  consider the bi-partite system $AB$.
The whole system evolves under  Hamiltonian (\ref{HXY}) which   can be separated into three parts:
\begin{eqnarray}\label{H}
H=H_A+H_B+H_{AB} ,
\end{eqnarray}
where $H_A$ and $H_B$  are the Hamiltonians governing the dynamics inside of, respectively, the subsystems $A$ and $B$, 
while $H_{AB}$ is the Hamiltonians describing the interactions between these  subsystems. 
Thus,
\begin{eqnarray}
\rho^{AB}(t)= e^{-i H t} \rho_0^{AB} e^{i H t},
\end{eqnarray}
where $\rho_0^{AB}$ is an initial density matrix of the system $AB$.  
Suppose that at some time instant $t_0$ we switch off the interaction, i.e. $H_{AB}|_{t>t_0}=0$, 
so that each  subsystem $A$ and $B$  evolves under its Hamiltonian, respectively, $H_A$ and $H_B$:
\begin{eqnarray}\label{ev2}
\rho^{AB}(t)= \left( e^{-i H_A t} \otimes e^{-i H_B t}\right)  \rho^{AB}(t_0) \left( e^{i H_A t} \otimes e^{i H_B t} \right), \;\; t>t_0.
\end{eqnarray}
Suppose that we would like to store the state of the subsystem $B$,
\begin{eqnarray}
\rho^B(t)=Tr_A\rho^{AB}(t),
\end{eqnarray}
during  some time interval $\Delta t$, so that 
\begin{eqnarray}
\rho^B(t_0+\Delta t)=\rho^B(t_0) .
\end{eqnarray}
{ We require also that the subsystem $A$ continues evolution under the Hamiltonian $H_A$ during the time interval $\Delta t$. Such a ''splitting'' of evolution is possible only if the hyromagnetic ratios of particles composing subsystems $A$ and $B$ are different.   In this case we can} assume that  the evolution of the subsystem  $B$ can be  organized according to the protocol
 described in Sec.\ref{Section:TR1} { without destroying the evolution of the subsystem $A$}, 
 i.e., we store the state using the identity-evolution operator
 $V$ {  given  by
 one of forms (\ref{V},\ref{tau}),   (\ref{ev21}),   or 
 (\ref{ev}).}
We have 
 \begin{eqnarray}
 \rho^B(t_0+n T)= V^n \rho^B(t_0) (V^+)^n \equiv  \rho^B(t_0),
 \end{eqnarray}
 where we take $\Delta t = n T$.
Then, at the time instant $t=t_0+ n T$, we switch off the { state-storing process},
again switch on the interaction between the subsystems $A$ and $B$ and let the whole system evolve under the Hamiltonian $H$ (\ref{H}). Thus, the state of the subsystem $B$ { has been} frozen during the time interval $\Delta t$.

 \subsection{State transfer and storage}
 \label{Section:sttr}
 Now we split the  subsystem $A$ into  the sender $S$ and the transmission line $TL$, therewith the dimensionality of the sender equals the dimensionality of the subsystem $B$ which is called  a receiver. 
 We consider  the tensor product initial state, 
 \begin{eqnarray}
 \rho(0) = \rho^S(0)\otimes \rho^{TL}(0) \otimes \rho^{B}(0),
 \end{eqnarray}
 where both subsystems $TL$ and $B$  are in the ground state { and $\rho^S(0)$ is the state we need to transfer to the subsystem $B$ and store therein}. 
 We also require that the whole system is engineered for the perfect state transfer \cite{Bose,CDEL}, so that,
 at some time instant  $t=t_0$, we have the subsystems $\rho^S(t_0)$ and  $\rho^{TL}(t_0)$ in the ground states, while 
 $\rho^{B}(t_0)$ is equivalent to $\rho^{S}(0)$ up to { the reordering the basis.}  
 
 At  $t=t_0$, we switch off the interaction between the subsystems $TL$ and $B$. After that,  the subsystem $S-TL$  
 can be used for some different purpose, while the state of $B$ is stored via the identity-evolution  operator $V$ {  given  by
 one of forms (\ref{V},\ref{tau}),   (\ref{ev21}),   or 
 (\ref{ev}).} { Notice that to provide the independency of the governing Hamiltonians for subsystems $B$ and $S-TL$ these subsystems must consist  of spins with 
 different 
 hyromagnetic ratios, similar to Sec.\ref{Section:frozen}. }
 Next, we provide the ground state of the subsystem $S-TL$ at $t=t_0+n T$, $n\in \NN$, then switch  off the { state-storing process} 
 and switch on { the interaction between the subsystems $TL$ and $B$}, therefore the system continue evolution under the Hamiltonian 
 $H$  (\ref{H}).
 Next, since the spin system  { admits} the perfect state transfer, the state of $B$ will be transferred back to $S$ at the time instant
 $t=2 t_0 + n T $. {  This effect can be used in  delay lines.}
 
 \subsection{Control of interaction between subsystems { $A$ and $B$}} 
 
 We see in the above  examples   (Secs.\ref{Section:frozen} and \ref{Section:sttr} )
 that  breaking the 
 interaction between subsystems is  an important step in the  state-conservation protocol. 
 The interaction between the subsystems $A$ and $B$ can be simply broken in the case of nearest neighbor interactions { as follows}.
 Suppose that the subsystems $A$ and $B$ consist of  {  particles with the gyromagnetic ratios, respectively, $\gamma_1$ and $\gamma_2$,
 and there is an impurity particle between $A$ and $B$, so that $ A$ interacts with $B$ through this impurity.  We require, that the gyromagnetic ratio
 $\gamma_3$ of the impurity is bigger then $\gamma_1$ and $\gamma_2$.} Then, applying the resonance (for the impurity) magnetic field in the direction perpendicular to the $z$-directed strong external field, we  cause { fast} precessing  of the  impurity's spin so that its average interaction with neighbors vanishes and, as a consequence, the interaction between $ A$ and $B$ breaks.   Removing the above perpendicular magnetic field  we 
 return the interaction between $ A$ and $B$.

\section{Conclusion}
\label{Section:conclusion}
The proposed protocol is an effective method of  storage of the information encoded into a multi-qubit state. It has no restrictions on the size of the  stored state and this state can be well conserved { up to the restrictions connected with the first non-vanishing term   of the  
  average Hamiltonian theory \cite{HW}} used in constructing the identity evolution operator $V$ (\ref{ev21}) or (\ref{ev}). { The  protocol 
  described in Sec.\ref{Section:TR1} is prepared for a spin system with particles having the same hyromagnetic ratios (homonuclear systems). The generalization to particles with different 
  hyromagnetic ratios (heteronuclear systems) is also possible. Notice that the state  in examples of Sec.\ref{Section:application} is also stored in the homonuclear subsystem (subsystem $B$) while another subsystem consisting of   particles having different hyromagnetic ratio (subsystem $A$) is used for different purpose, for instance, to transfer the needed state to the subsystem $B$. } 
  
  { We shall notice that the time of spin-lattice relaxation in the crystals of calcium fluorapatite and  calcium hydroxyapatite 
  is much longer then the time of a particular quantum operation. Therefore we neglect the interaction with environment in this paper, although spin-lattice relaxation, in general,  makes restrictions on the storage time-intervals.}
  Another problem to study is constructing the multi-block memory allowing us to store a set of multi-qubit states in 
a register of proper dimensionality. 

The work is partially supported by the Russian Foundation  for basic research (Grants 16-03-00056 and 15-07-07928).

\end{document}